\documentclass[preprint,11pt,authoryear]{elsarticle}
\usepackage{amssymb}
\usepackage{lipsum}

\journal{Icarus}
\usepackage{lineno}
\usepackage{makecell}
\usepackage[colorlinks=true, linkcolor=blue, citecolor=blue, urlcolor=blue]{hyperref}
\usepackage{academicons}
\usepackage{xcolor}
\usepackage{url}
\usepackage[margin=3.0cm]{geometry}

\begin{document}
\begin{frontmatter}
\title{Spectral Mixture Modeling with Laboratory Near-Infrared Data II: Effects of Grain Size and Implications for Europa}

\author[first]{A. Emran}
\affiliation[first]{organization={NASA Jet Propulsion Laboratory},
            addressline={California Institute of Technology}, 
            city={Pasadena},
            postcode={91109}, 
            state={CA},
            country={USA}}
\ead{al.emran@jpl.nasa.gov || al.emraan@gmail.com}
\begin{abstract}
Spectral analysis using linear mixture (LM) and radiative transfer–based (RT) intimate mixture modeling based on Hapke theory at near-infrared wavelengths are applied to estimate the abundance of surface materials on Europa. Previously, \cite{emran2025paper1} compared these approaches against the laboratory spectra of H\textsubscript{2}O ice and H\textsubscript{2}SO\textsubscript{4}$\cdot$8H\textsubscript{2}O mixtures with $\sim$100 $\mu$m grains. Here, the effect of particle size on spectral modeling accuracy was assessed using laboratory spectra of H\textsubscript{2}O ice mixtures with small ($\sim$70 $\mu$m spherical) and coarse ($\sim$1 mm irregular) grains, measured over the $\sim$1.2-2.5 $\mu$m wavelength range at 100 K and 120 K \citep{stephan2021vis}. Modeled abundance estimates at both temperatures show consistent trends across all mixing ratios, with only minor temperature-dependent variations. The discrepancy in abundance estimates from both LM and RT models remains within $\pm$10\% across all mixtures, with the error reduced to $\pm$5\% when fine grains dominate. Across all mixtures, the average difference between RT- and LM-derived abundance estimates remains within $\pm$2\% for mixtures containing both small and large grains. In contrast, mixtures composed solely of smaller grains render larger deviations between the models, with RT producing more accurate estimates \citep{emran2025paper1}— indicating that the presence of coarse H\textsubscript{2}O ice grains minimizes abundance differences between LM and RT modeling. Thus, I posit that Hapke-based RT modeling is the preferred spectral modeling approach— regardless of grain size or compositional mixture— for constraining Europa’s surface composition. Nonetheless, LM modeling remains a reliable approach for compositional analysis of terrains containing H\textsubscript{2}O ice with $\sim$mm-sized grains.
\end{abstract}

\end{frontmatter}

\section{Introduction}
\label{introduction}
Water (H\textsubscript{2}O) ice, a major component of Jupiter’s moon Europa, has been detected through both ground- and space-based observations \citep{moroz1966spectra, pilcher1972galilean,anderson1997europa,mccord1998salts, carlson2009europa}. Accurate estimates of H\textsubscript{2}O ice characteristics on planetary bodies are essential for understanding volatile transport, geological and morphological evolution, and surface dynamics of icy Solar System bodies \citep{baragiola2003water,prockter2005ice,kofman2019refractive,stephan2021vis}. Over the past decades, the quantitative estimation of surface materials— both ice and non-ice— on Europa has been conducted using spectral analysis techniques, primarily through linear mixture (LM) modeling \citep[e.g.,][]{shirley2010europa,dalton2012europa, dalton2013exogenic,shirley2016europa, prockter2017surface, berdis2022europa, davis2024pwyll} and radiative transfer (RT) theory–based intimate mixture modeling \citep[e.g.,][]{carlson2005distribution, mishra2021comprehensive, mermy2023selection}.

Linear mixture modeling (also referred to as the areal or macroscopic mixture approach) assumes that each material in the surface mixture is spatially segregated, occupying distinct geographic patches \citep[e.g.,][]{adams1986spectral,adams1993imaging,cruikshank1993ices,clark1998multispectral,dalton2007linear,dalton2012europa, manolakis2016}. Under this assumption, the reflectance of the mixture is represented as a linear combination of the reflectance spectra of its individual components \citep[e.g.,][]{heinz2001fully,stack2015modeling, shimabukuro2018linear,emran&stack2025}. In contrast, intimate mixture modeling based on radiative transfer theory considers that materials are physically intermixed at fine scales \citep[e.g.,][]{hapke1981bidirectional,mustard1987quantitative,poulet2004nonlinear, li2011radiative, li2015estimating}. This approach accounts for complex light scattering processes, as the reflectance spectra of an intimate mixture show nonlinear combinations of the spectra of the individual constituents \citep[e.g.,][]{nash1974spectral,johnson1983semiempirical,clark1984spectral}. Radiative transfer modeling, particularly based on Hapke theory \citep{hapke1981bidirectional}, has been widely applied for estimating surface properties of icy bodies \citep[refer to][for comprehensive references]{emran&chevrier2023, khuller2025quantitative}.

Both LM and Hapke-based \citep{hapke1981bidirectional, hapke2002bidirectional, hapke2012theory} RT modeling have successfully been applied to estimate the quantitative abundance of surface compositions on Europa using telescopic and spacecraft observations at near-infrared (NIR; $\sim$1 - 2.5 $\mu$m) wavelengths. However, the accuracy of these modeling approaches remains poorly constrained, despite the fact that a reliable understanding of Europa’s surface requires accurate compositional estimates. \cite{shirley2016europa} initiated efforts to comparing spectral LM and RT modeling using spacecraft observations from the Galileo Near-Infrared Mapping Spectrometer \citep[NIMS;][]{carlson1992near}. Their analysis showed that both models produced broadly similar abundance estimates of water ice and sulfuric acid (H\textsubscript{2}SO\textsubscript{4}$\cdot$\textit{n}H\textsubscript{2}O)— with a linear correlation coefficient of $\ge$ 0.98— across the different regions of Europa \citep[refer to ][for details]{shirley2016europa}. 

More recently, \citet{emran2025paper1} validated the accuracy of LM and Hapke-based RT modeling using laboratory spectra of H\textsubscript{2}O ice and sulfuric acid octahydrate (SAO; H\textsubscript{2}SO\textsubscript{4}$\cdot$8H\textsubscript{2}O) mixtures at varying mixing ratios, with particle size of $\sim$100 $\mu$m for both components \citep{hayes2025insights}. The study found that Hapke-based RT modeling is generally preferred for constraining Europa’s surface composition, although LM also produced viable results under specific compositional regimes for the same grain sizes in the mixtures. This conclusion is based on the observation that, for these small grain sizes, Hapke-based RT modeling estimated the true laboratory abundances more accurately than LM modeling. The improved abundance prediction by RT modeling can be attributed to its ability to account for the complex interaction of photon with particulate media and to incorporate multiple-scattering effects inherent to intimate mixtures \citep[e.g.,][]{mustard1987quantitative, mustard1989photometric} — a critical factor when particle sizes (\textit{d}) are much smaller than the photon penetration depth (\textit{$\delta$}) at NIR wavelengths. Under such conditions (\textit{d} \(<\)\(<\) \textit{$\delta$}), photons penetrate deeper into the granular media, interact with multiple grains, leading to significant multiple-scattering effects of light \citep{hapke1981bidirectional, hapke2012theory}. Intimate mixture modeling incorporates these effects through implementation of radiative transfer theory \citep{chandrasekhar2013radiative}, enabling more accurate abundance estimates, whereas linear mixture modeling does not account for the complex interaction of light with particulate media, resulting in reduced accuracy in abundance estimates of H\textsubscript{2}O ice and H\textsubscript{2}SO\textsubscript{4}$\cdot$8H\textsubscript{2}O mixtures with smaller grains. 

However, \citet{emran2025paper1} focused exclusively on mixtures with a single grain size ($\sim$100 $\mu$m) and did not assess the influence of particle size on modeling accuracy— particularly for $\sim$mm-sized grains in the mixture, as observed in many regions across Europa’s surface \citep{ligier2016vlt, king2022compositional, emran2025nh}. Grain size has a significant influence on the spectral features, such as depth of characteristic absorption band \citep[e.g.,][]{mastrapa2008optical, mastrapa2009optical} and, consequently, on the resulting spectral modeling and derived abundance estimates of H\textsubscript{2}O ice. In this study, I assess the accuracy of LM and Hapke-based RT modeling using laboratory spectra of H\textsubscript{2}O ice mixtures with small ($\sim$70 $\mu$m) and coarse ($\sim$1 mm) grains measured at 100 K and 120 K temperatures \citep{stephan2021vis}. This complementary investigation provides a framework for improving quantitative constraints on compositional analysis of Europa’s surface and is directly relevant to upcoming spectral analyses from JUICE’s Moons and Jupiter Imaging Spectrometer \citep[MAJIS;][]{poulet2024moons} and Europa Clipper’s Mapping Imaging Spectrometer for Europa \citep[MISE;][]{blaney2024mapping}. 

Previous studies \citep[e.g.,][]{hansen2009calculation,emran&chevrier2022, emran&chevrier2023} have investigated discrepancies in grain size estimations from NIR spectra by calculating and comparing single-scattering albedos for ices relevant to icy bodies in the outer Solar System, using Mie theory \citep{mie1908} and two approximate versions of Hapke's formulations \citep{hapke1981bidirectional, hapke1993}. In these studies, optical constants (\textit{n }and \textit{k}) of pure ices were used to compute single-scattering albedos from Hapke approximations to the Mie spectra, allowing them to determine the best-fit particle sizes corresponding to those albedo spectra of pure ices. Another relevant study by \citet{khuller2025quantitative} compared spectral albedos and estimated best-fit particle sizes for well-constrained reflectance spectra of H\textsubscript{2}O ice \citep{dadic2013effects} using widely used radiative transfer models, including the delta-Eddington \citep{joseph1976delta}, Hapke \citep{hapke1981bidirectional}, and Shkuratov \citep{shkuratov1999model} formulations. In contrast, the present study fundamentally differs from these works — it estimates the abundance of each component directly from measured binary mixture spectra and evaluates the accuracy of linear and intimate mixture modeling in reproducing the true abundances. More specifically, the influence of grain sizes on scattering properties, spectral modeling, and the resulting abundance estimates is explored with the present study in the context of mixtures containing both small and large grains — relevant to Europa.

\section{Observation and Methods}
\subsection{Laboratory spectra}
\begin{figure*}
	\centering 
	\includegraphics[width=1.\textwidth, angle=0]{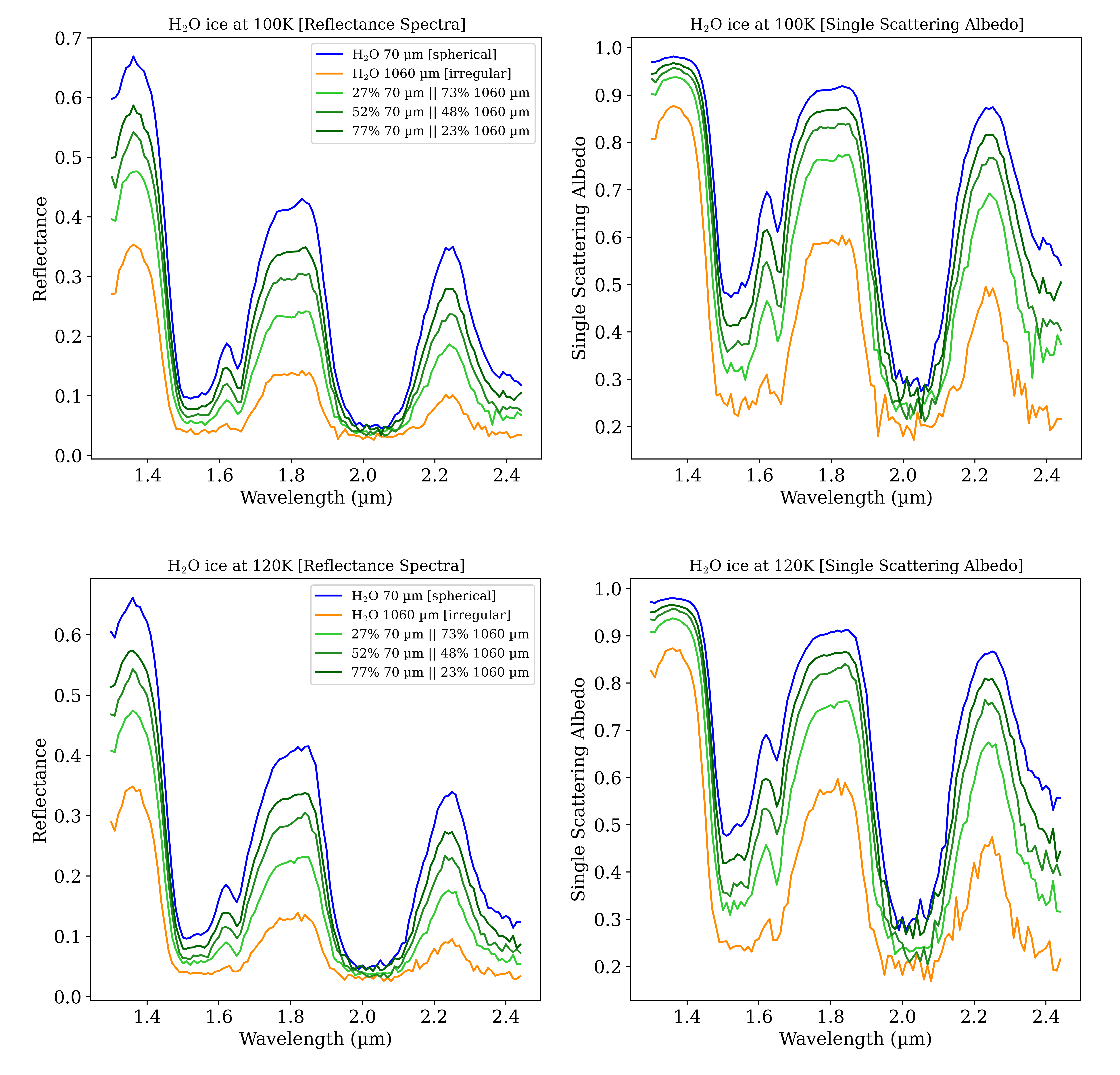}	
	\caption{\textit{Left panel:} Laboratory spectra of H\textsubscript{2}O ice at $\sim$70 $\mu$m spherical and $\sim$1060 $\mu$m irregular grains and their mixtures at different mixing ratios at the temperature of 100K (upper row) and 120K (lower row). The reflectance spectra were collected from \cite{stephan2021vis}. \textit{Right panel:} Single scattering albedo spectra calculated from the reflectance spectra (on the corresponding left panels) using the \cite{hapke1981bidirectional} model.} 
	\label{fig1}%
\end{figure*}

The spectra of H\textsubscript{2}O ice mixtures with particle sizes of $\sim$70 $\mu$m\footnote{Measured size of 70 $\pm$ 30 $\mu$m} (small grains) and $\sim$1060 $\mu$m\footnote{Measured size of 1060 $\pm$ 60 $\mu$m}  ($\sim$1mm; large grains) were measured at mixing ratios (\%wt) of 27\% : 73\%, 52\% : 48\%, and 77\% : 23\%, respectively \citep[Fig. \ref{fig1};][]{stephan2021vis}. The small grains ($\sim$70 $\mu$m) were spherical in shape, while the large grains ($\sim$1060 $\mu$m) were irregular. Although the laboratory reflectance spectra were collected at a range of temperatures, for this study, I used only the reflectance spectra measured at 100 K and 120 K, conditions relevant to Europa \citep[e.g.,][]{spencer1999temperatures,ashkenazy2019surface, carlson2009europa}. While Europa's surface shows a broad temperature range, as the global mean annual surface temperatures of 90 K \citep[e.g.,][]{ashkenazy2019surface}, the selected temperatures of 100 K and 120 K can represent the daytime surface conditions  \citep[e.g.,][]{spencer1999temperatures, grundy1999near,carlson2009europa, rathbun2010galileo}. Moreover, laboratory measurements across the temperature range of 70–150 K indicate that the depths of major H\textsubscript{2}O ice absorption bands remain relatively stable, showing only minor variations \citep[e.g.,][]{stephan2021vis}. Thus, the measurements at 100 K and 120 K can be considered representative of Europa’s surface temperature conditions, as used in this study. Spectral measurements of H\textsubscript{2}O ices were done with an incidence angle (\textit{i}) of 0° and an emission angle (\textit{e}) of 30°— resulting in an equivalent phase angle (\textit{g}) of 30° —  and a spectral sampling of 0.010 $\mu$m over the 1.0–3.6 $\mu$m wavelength range \citep{stephan2021vis}. The $\sim$70 $\mu$m spherical and $\sim$1060 $\mu$m irregular particle spectra were used as endmembers for both LM and RT spectral modeling. Note that the laboratory measurements have an absolute radiometric accuracy of $\sim$3\%. For details of the experimental setup and measurement procedures, see \cite{stephan2021vis}. 

\subsection{LM and RT modeling}

The reflectance of a mixture (\textit{r}\textit{\textsubscript{m}}) using LM modeling can be written as \citep[e.g.,][]{emran2023surface}:

\begin{equation}
    r_m = \sum_{i=1}^{k} f_i \cdot r_i ~ ;  0 \leq f_i \leq 1 ~;\sum_{i=1}^{k} f_i = 1
    \label{eq1}
\end{equation}

where \textit{f}\textit{\textsubscript{i}} is the fractional abundance of the \textit{i\textsuperscript{th}}  endmember with reflectance \textit{r}\textit{\textsubscript{i}}  \citep[e.g.,][]{dalton2012europa, emran2021thermophysical, davis2024pwyll}.

The reflectance spectra of an intimate mixture of minerals or particulate media show systematic but complex, non-linear scattering behavior, resulting in a non-linear combination of reflectance spectra of constituent components  \citep[e.g.,][]{nash1974spectral,mustard1989photometric, mustard2019theory}. However, when reflectance is converted to single-scattering albedo using radiative transfer theory, the scattering behavior of a mixture can be approximated as a linear combination of single-scattering albedo spectra of each component \citep[e.g.,][]{johnson1983semiempirical, mustard1987quantitative}. Thus, for RT-based intimate mixture modeling, reflectance or radiance coefficient (\textit{r}) is first converted to  single scattering albedo (\textit{w}) using Hapke theory \citep{hapke1981bidirectional}:

\begin{equation}
r(i, e, g) = \frac{w}{4} \cdot \frac{1}{\mu_e + \mu_i} 
\left\{ \left[1 + B(g)\right] P(g) + H(\mu_e) H(\mu_i) - 1 \right\}
\label{eq:2}
\end{equation}

where \textit{$\mu$\textsubscript{i}} and \textit{$\mu$\textsubscript{e}} are the cosines of incidence angle (\textit{i}) and emission angle (\textit{e}), respectively; \textit{B}(\textit{g}) and \textit{P}(\textit{g}) are the backscattering and the single-particle phase functions, respectively; and \textit{H($\mu$\textsubscript{x})} represents \cite{chandrasekhar2013radiative}'s multiple scattering functions. Following the approach of \cite{emran2025paper1}, reflectance spectra were converted to single-scattering albedo (cf. right column of Fig. \ref{fig1}). In this instance, I consider no backscattering (\textit{B}(\textit{g}) = 0) and an isotropic phase function (\textit{P(g) = 1}), such that the Eq. \ref{eq:2} can be simplified as \citep[e.g.,][]{mustard1987quantitative,mustard1989photometric}:

\begin{equation}
r(i, e, g) \approx \frac{w}{4} \cdot \frac{1}{\mu_e + \mu_i} \cdot H(\mu_e) H(\mu_i)
\label{eq:3}
\end{equation}

Note that the assumption of an isotropic phase function (\textit{P(g)}=1) is a simplification, as scattering in particulate media can be non-isotropic \citep[e.g.,][]{hapke2012theory}. In Hapke-type models, the phase function, single-scattering albedo, and grain size are strongly coupled —  an inconsistent assumption regarding the phase function can be compensated by adjustments in other parameters to maintain a good spectral fit. Thus, the parameters derived using this simplified configuration may be interpreted as apparent (model-dependent) quantities rather than absolute physical parameters. However, the potential for such parameter degeneracy is minimized in this study because the particle sizes are not retrieved as free parameters but are pre-constrained by the known laboratory sample preparations \citep[$\sim$70 $\mu$m and $\sim$1060 $\mu$m;][]{stephan2021vis}. Furthermore, the use of an isotropic approximation is justified for estimating compositional abundances, as it has been validated to provide accuracy within a few percent (\%) for the specific measurement geometries (intermediate phase angles of 15°–40°) and known particle sizes used in this study \citep{mustard1987quantitative, mustard1989photometric}.

The validation of this simplified reflectance model (Eq. \ref{eq:3}) has been tested using laboratory spectra of mineral mixtures, and the results show (as mentioned above) that it provides abundance estimates within a few percent when grain sizes are well constrained — similar to this study \citep{mustard1987quantitative, mustard1989photometric, emran2025paper1}. Particularly, the abundance estimates derived from the simplified approach (Eq. \ref{eq:3}) have been validated under the assumption of isotropic scattering of light for particulate media in intimate mixtures at phase angles of 15°–40°, where backscattering is considered negligible (\textit{B(g)} \(<\)\(<\) 1) — phase angles comparable to those used in this study and often used by imaging spectrometers
 \citep{mustard1987quantitative, mustard1989photometric}. Thus, this simplified version of the reflectance model was adopted here for the RT modeling \citep{lapotre2017probabilistic}. The multiple scattering \textit{H}-functions were then calculated as \citep{hapke2002bidirectional}:

\begin{equation}
H(x) \approx \left[ 1 - wx \left( r_o + \frac{1 - 2 r_o x}{2} \ln\left( \frac{1 + x}{x} \right) \right)^{-1} \right]
\label{eq:4}
\end{equation}

\begin{equation}
r_o = \frac{1 - \gamma}{1 + \gamma}
\label{eq:5}
\end{equation}

\begin{equation}
\gamma = \sqrt{1 - w}
\label{eq:6}
\end{equation}

where \textit{x} is cosine of the incidence ($\mu$\textsubscript{i}) or emission ($\mu$\textsubscript{e}) angles, \textit{r\textsubscript{o}} is the bihemispherical reflectance for isotropic scatterers, and $\gamma$  is the albedo factor. Note that detail of the equations can also be found in relevant literature \citep[e.g.,][]{hapke1981bidirectional,hapke1993, hapke2001space, hapke2002bidirectional,hapke2008bidirectional,hapke2012theory}. In RT-based intimate mixture modeling, the single scattering albedo of a mixture (\textit{w\textsubscript{m}}) can be expressed as \citep[e.g.,][]{johnson1983semiempirical,stack2015modeling, goudge2015integrating, lapotre2017probabilistic}:

\begin{equation}
w_m = \sum_{k=1}^{i} f_k \cdot w_k
\label{eq:7}
\end{equation}

where \textit{w\textsubscript{k}} and \textit{f\textsubscript{k}} are the single scattering albedo and relative fractional (geometric) cross section of the \textit{k\textsuperscript{th}} endmember, respectively. The parameter \textit{f\textsubscript{k}} is function of the mass fraction, density, and particle diameter of corresponding \textit{k\textsuperscript{th}} endmember \citep[e.g.,][]{hapke1981bidirectional,mustard1987quantitative, mustard1989photometric}. Particle size has a major influence on the spectral characteristics \citep[e.g.,][]{stephan2021vis}, however, particle shape can also affect the scattering properties \citep[e.g.,][]{shkuratov2005light}. For particles of similar sizes and densities, relative fractional cross section can approximate to the mass fractional abundance, specifically when particle sizes in a mixture are known \citep{mustard1987quantitative}. However, the larger ($\sim$1060 $\mu$m) H\textsubscript{2}O ice grains in this study were irregularly shaped, resulting in a lower average cross-sectional scattering area— reduced mean photon path length or "effective diameter" (\textit{$\langle D \rangle$}) — compared to spherical grains \citep[e.g.,][]{hulst1981light, grenfell1999representation,bohren2008absorption, hapke2012theory}. Thus, the relative fractional cross section for the $\sim$1060 $\mu$m grains was scaled by a shape factor of 2/3, as irregular or rough-surfaced particles typically have an "effective diameter" that range from 2/10 to 9/10 of the particle diameter \citep[e.g.,][]{shkuratov2005light}. Recently, \citet{berdis2025near} also used the same scaling factor of 2/3 for the appropriate grain size in the spectral modeling of laboratory mixtures of water ice and epsomite — relevant to surface composition of Europa. The fractional abundances were adjusted accordingly. Lastly, the Markov Chain Monte Carlo (MCMC) technique \citep[e.g.,][]{hogg2018data} was implemented using the \textit{emcee} Python package \citep{foreman2013emcee} to estimate model abundances from 1,000 iterations. The priori parameter values were estimated close to the expected true values, calculated using a SciPy \citep{SciPy2020} optimization module prior to running the MCMC routine. In this analysis, a burn-in period of 100 steps was applied to the MCMC chain. The model-derived abundances were then obtained as the mean ± 1$\sigma$ uncertainties from the posterior distributions.

\section{Results}
\begin{figure*}
	\centering 
	\includegraphics[width=1.\textwidth, angle=0]{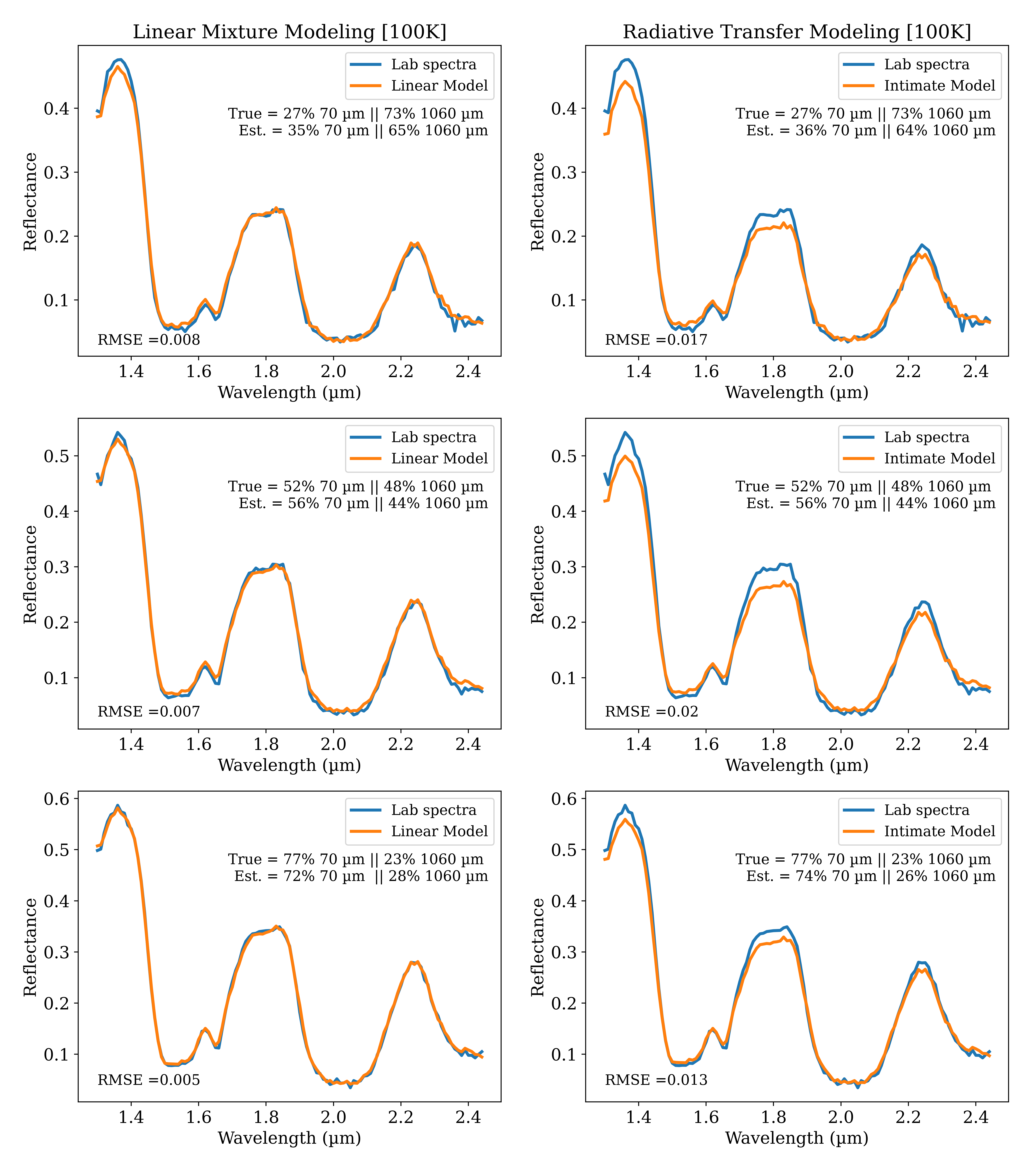}	
	\caption{Comparison of LM (left panel) and RT (right panel) modeling results for H\textsubscript{2}O ice reflectance measurements at 100K. The blue curves represent laboratory reflectance spectra of H\textsubscript{2}O ice mixtures of $\sim$70 $\mu$m spherical and $\sim$1060 $\mu$m irregular grains at different ratios \citep{stephan2021vis}, while the orange curves represent modeled spectra from LM or RT modeling. Listed are the laboratory and model-derived mean abundances of H\textsubscript{2}O grains, along with the root mean square error (RMSE) of each spectral fit.} 
	\label{fig2}%
\end{figure*}

\begin{figure*}
	\centering 
	\includegraphics[width=1.\textwidth, angle=0]{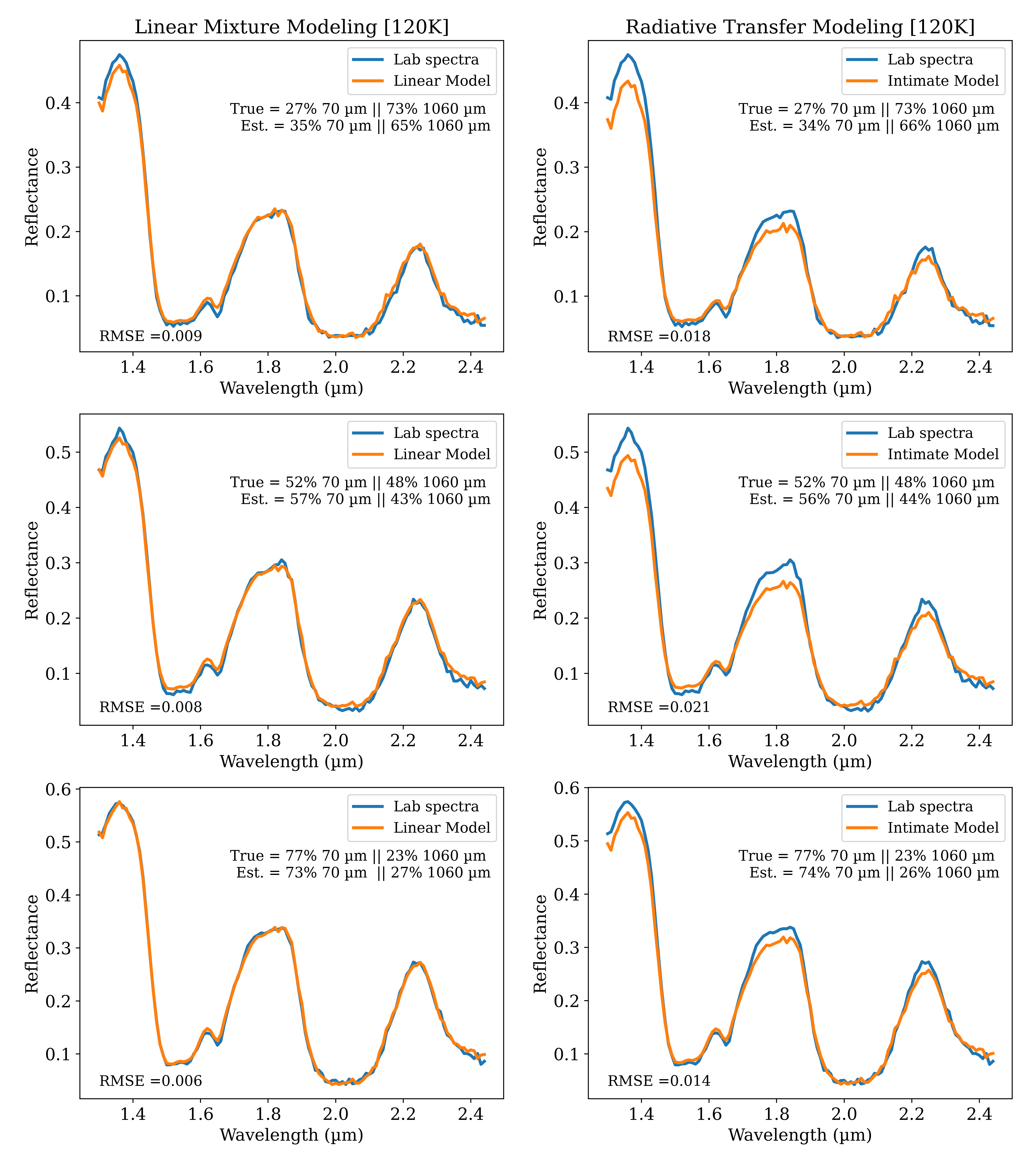}	
	\caption{Comparison of LM (left panel) and RT (right panel) modeling results for H\textsubscript{2}O ice reflectance measurements at 120K. The blue curves represent laboratory reflectance spectra of H\textsubscript{2}O ice mixtures of $\sim$70 $\mu$m spherical and $\sim$1060 $\mu$m irregular grains at different ratios \citep{stephan2021vis}, while the orange curves represent modeled spectra from LM or RT modeling. Listed are the laboratory and model-derived mean abundances of H\textsubscript{2}O grains, along with the root mean square error (RMSE) of each spectral fit.} 
	\label{fig3}%
\end{figure*}

\begin{figure*}
	\centering 
	\includegraphics[width=1.\textwidth, angle=0]{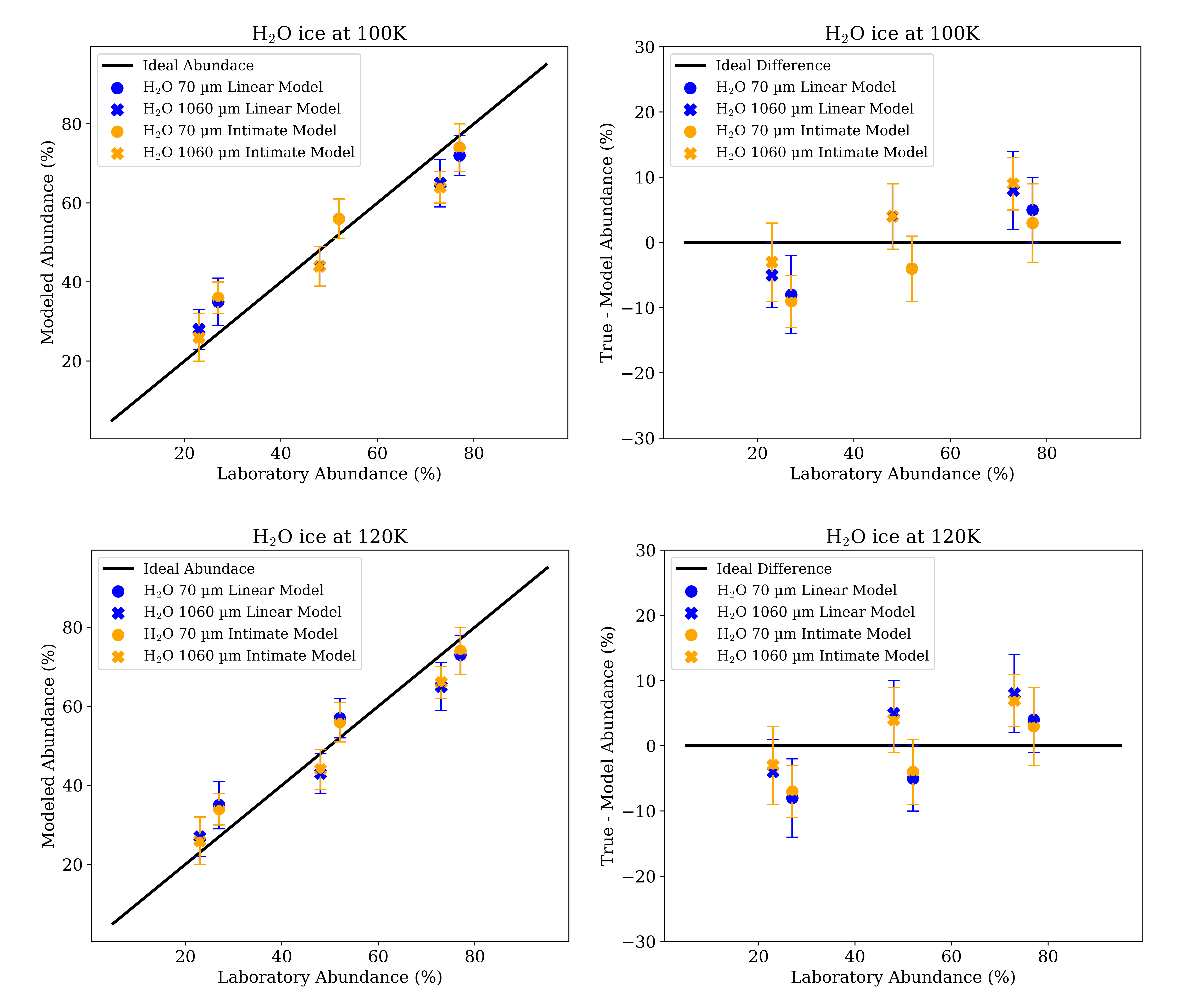}	
	\caption{Modeled abundances (mean ± 1$\sigma$) of H\textsubscript{2}O ice mixtures with $\sim$70 $\mu$m and $\sim$1060 $\mu$m grains at 100K (upper panel) and 120K (lower panel) using LM and RT modeling. \textit{Left column}: Estimated abundances compared to expected laboratory values. \textit{Right column}: Discrepancies between laboratory and modeled abundances (True – Estimated) for each grain size using LM and RT modeling. Blue and orange markers indicate estimates from LM and RT modeling, respectively; circles and crosses denote $\sim$70 $\mu$m and $\sim$1060 $\mu$m grain abundances. Note that the model-derived abundances (mean ± 1$\sigma$ uncertainties) were obtained from the posterior distributions generated using the MCMC routine. Thus, the error bars represent the 1$\sigma$ standard deviations of the posterior abundance distributions produced by the MCMC simulations.}
	\label{fig4}%
\end{figure*}

Figs. \ref{fig2} and \ref{fig3} show the results of LM and RT modeling for the reflectance spectra of H\textsubscript{2}O ice mixtures at 100K and 120K, respectively, along with the corresponding estimated abundances of the $\sim$70 $\mu$m (small) and $\sim$1060 $\mu$m (large) grain sizes. Each row in the subplots of Figs. (\ref{fig2} - \ref{fig3}) represents a different mixing ratio, with the modeled root mean square error (RMSE) calculated for both modeling techniques. The mean estimated abundances and associated $\pm 1\sigma$ uncertainties by LM and RT modeling for both grain sizes at different mixing ratios are summarized in Table \ref{tab1} (100K) and Table \ref{tab2} (120K). 

For the laboratory spectra of H\textsubscript{2}O with a 27\% ($\sim$70 $\mu$m) : 73\% ($\sim$1060 $\mu$m) mixture, the LM model estimates abundances of 35\% small grains and 65\% large grains at both 100K and 120K (Tables \ref{tab1} and \ref{tab2}). This result indicates that temperature differences from 100K to 120K do not significantly affect the LM-derived abundances. In contrast, RT modeling results in similar but slightly varying estimates, with 36\% $\&$ 34\% small grains, and 64\% $\&$ 66\% large grains for the 100K and 120K cases, respectively. At both temperatures, however, the RMSE of the spectral fit is lower for LM than for RT modeling (Figs. \ref{fig2} and \ref{fig3}). Nonetheless, both models reproduce abundances within $\pm 10$\% of the true laboratory values for this specific mixture of small and large H\textsubscript{2}O grains (Fig. \ref{fig4}).

As the smaller grain ($\sim$70 $\mu$m) abundance of H\textsubscript{2}O ice is 52\% in the mixture, both LM and RT modeling estimate an abundance of 56\% for the $\sim$70 $\mu$m grain at 100K (Table \ref{tab1}). However, at 120K, the LM estimate slightly increases to 57\%, while the RT estimate remains unchanged at 56\% (Table \ref{tab2}). At this mixture ratio, the modeled abundances are estimated within $\pm 5$\% of the true values by both LM and RT modeling at 100K and 120K (Fig. \ref{fig4}). Nonetheless, the RMSE value remains lower for LM than for RT modeling (Figs. \ref{fig2} and \ref{fig3}).

Lastly, when the small grain of H\textsubscript{2}O dominates in the mixture (77\% $\sim$70 $\mu$m : 23\% $\sim$1060 $\mu$m), the RT modeling abundance estimate is slightly better, with a consistent predicted value of 74\% for the small grain at both 100K and 120K (Tables \ref{tab1} and \ref{tab2}). By comparison, LM produces slightly lower abundances of 72\% and 73\% for the small grain at 100K and 120K, respectively. Nonetheless, the abundances estimated by both LM and RT are within $\pm 5$\% of the true values at both temperatures (Fig. \ref{fig4}). The RMSE value, however, remains lower for LM than for RT modeling (Figs. \ref{fig2} and \ref{fig3}).

\begin{table*}[ht]
\centering
\renewcommand{\arraystretch}{1.2}
\caption{Estimated abundances of $\sim$70 $\mu$m and $\sim$1060 $\mu$m H$_2$O ice at 100K using LM and RT modeling, reported fractional percentages (\%wt) as mean ± 1$\sigma$.}
\label{tab1}
\small
\begin{tabular}{lcccc}
\hline
\makecell{\textbf{Mixture abundance} \\ \textbf{\textbf{(H}\textbf{\textsubscript{2}}\textbf{O mixture \%)}}} &
\multicolumn{2}{c}{\makecell{\textbf{Linear mixture} \\ \textbf{abundance (\%)}}} &
\multicolumn{2}{c}{\makecell{\textbf{Intimate mixture} \\ \textbf{ abundance (\%)}}} \\
\cline{2-5}
 & 70 $\mu$m & 1060 $\mu$m  & 70 $\mu$m & 1060 $\mu$m  \\
\hline
27\% 70 $\mu$m + 73\% 1060 $\mu$m & 35 ± 6 & 65 ± 6 & 36 ± 4 & 64 ± 4 \\
52\% 70 $\mu$m + 48\% 1060 $\mu$m & 56 ± 5 & 44 ± 5 & 56 ± 5 & 44 ± 5 \\
77\% 70 $\mu$m + 23\% 1060 $\mu$m & 72 ± 5 & 28 ± 5 & 74 ± 6 & 26 ± 6 \\
\hline
\end{tabular}%
\end{table*}

\begin{table*}[ht]
\centering
\renewcommand{\arraystretch}{1.2}
\caption{Estimated abundances of $\sim$70 $\mu$m and $\sim$1060 $\mu$m H$_2$O ice at 120K using LM and RT modeling, reported fractional percentages (\%wt) as mean ± 1$\sigma$.}
\label{tab2}
\small
\begin{tabular}{lcccc}
\hline
\makecell{\textbf{Mixture abundance} \\ \textbf{\textbf{(H}\textbf{\textsubscript{2}}\textbf{O mixture \%)}}} &
\multicolumn{2}{c}{\makecell{\textbf{Linear mixture} \\ \textbf{abundance (\%)}}} &
\multicolumn{2}{c}{\makecell{\textbf{Intimate mixture} \\ \textbf{ abundance (\%)}}} \\
\cline{2-5}
 & 70 $\mu$m & 1060 $\mu$m  & 70 $\mu$m & 1060 $\mu$m  \\
\hline
27\% 70 $\mu$m + 73\% 1060 $\mu$m & 35 ± 6 & 65 ± 6 & 34 ± 4 & 66 ± 4 \\
52\% 70 $\mu$m + 48\% 1060 $\mu$m & 57 ± 5 & 43 ± 5 & 56 ± 5 & 44 ± 5 \\
77\% 70 $\mu$m + 23\% 1060 $\mu$m & 73 ± 5 & 27 ± 5 & 74 ± 6 & 26 ± 6 \\
\hline
\end{tabular}%
\end{table*}

\section{Discussion and implications for Europa}
Validation of spectral modeling approaches, such as linear mixture modeling and radiative transfer–based intimate mixture modeling, against laboratory measurements is crucial for accurately interpreting the surface composition, geology, and morphological dynamics of Europa \citep{shirley2016europa}. Toward this effort, \citet{emran2025paper1} compared LM and RT modeling against laboratory spectra of H\textsubscript{2}O ice and H\textsubscript{2}SO\textsubscript{4}$\cdot$8H\textsubscript{2}O mixtures with $\sim$100 $\mu$m grains \citep{hayes2025insights}, concluding that RT modeling provides better estimates than LM. Another recent study by \cite{berdis2025near} also compared linear and intimate mixture modeling using laboratory mixtures of water ice and sulfate salt, showing no substantial differences between the two spectral modeling approaches. As a complement to those works, the present study investigates the effect of grain size on scattering properties and consequently on spectral modeling by comparing the modeled abundances of H\textsubscript{2}O ice mixtures with varying proportions of small and large grains. The results of this study, provide important insights into the application of spectral modeling for estimating surface compositional abundances on Europa at near-infrared wavelengths ($\sim$ 1-2.5 $\mu$m).

\begin{figure*}
	\centering 
	\includegraphics[width=1.\textwidth, angle=0]{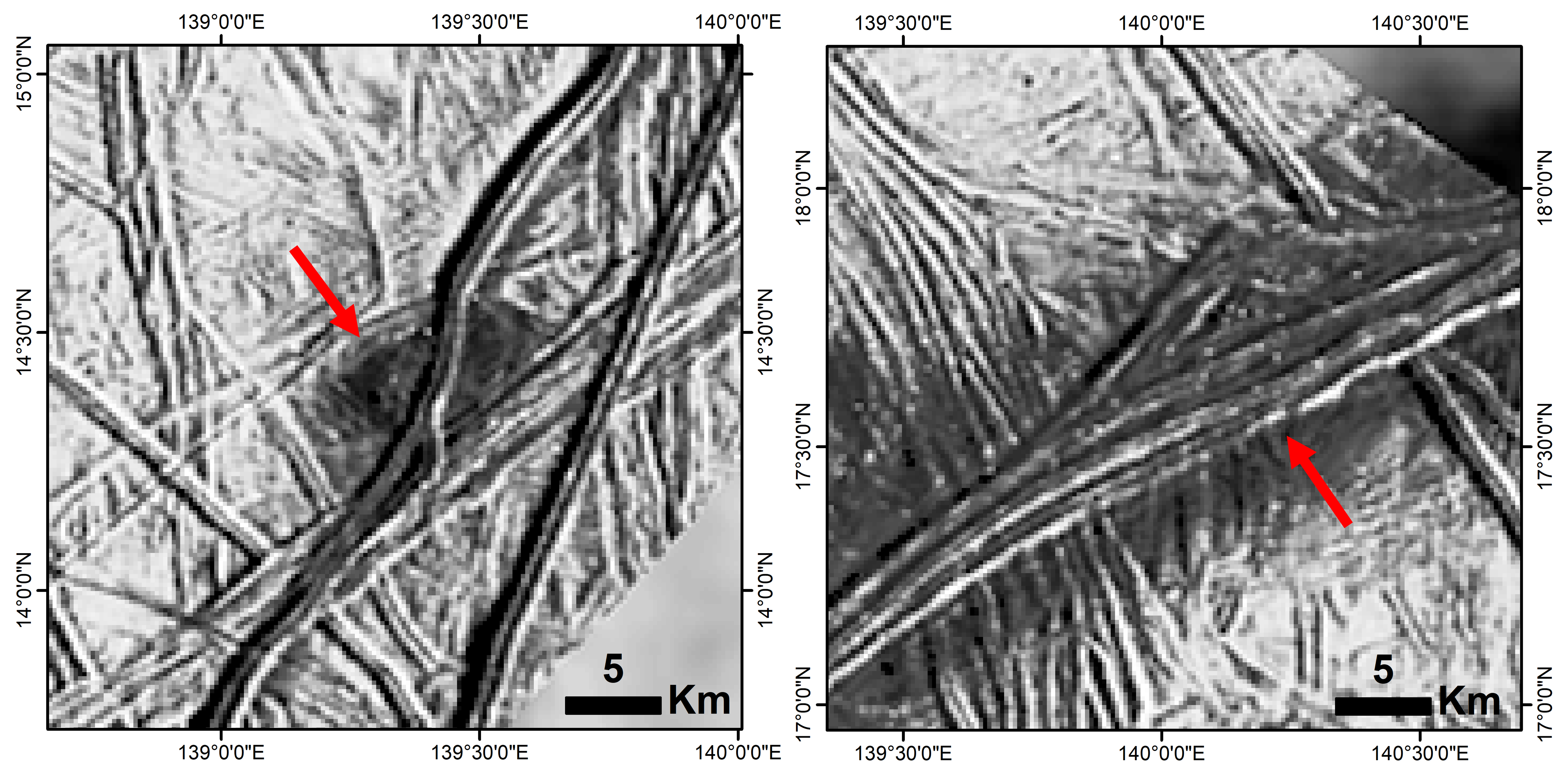}	
	\caption{\textit{Left panel:} A chaos feature on Europa (indicated by the red arrow), where ammonia (NH\textsubscript{3})-bearing compounds were modeled with $\sim$mm-sized H\textsubscript{2}O ice grains using near-infrared spectral data \citep{emran2025nh}. \textit{Right panel:} Linear feature, such as band terrain, on Europa (indicated by the red arrow). Both images are from the Galileo Solid State Imaging \citep[SSI;][]{belton1992galileo} instrument with a spatial resolution of $\sim$220 m/pixel, and adapted from \cite{malaska2024updated}.}
	\label{fig5}%
\end{figure*}

Across all mixing ratios, both LM and RT modeling provide abundance estimates within $\pm 10$\% of the true values for mixtures containing both small and large H\textsubscript{2}O ice grains. However, when the smaller grain dominates in the mixture, this uncertainty (error) decreases to within $\pm$5\% for both models. Although the RT estimated H\textsubscript{2}O ice abundances are slightly better across most mixture ratios, the difference between LM- and RT-derived estimates remains within $\pm$2\% for both small and large H\textsubscript{2}O ice grains. Thus, while RT is generally preferred for spectral modeling of surface composed solely of smaller grains regardless compositional mixture on Europa \citep{emran2025paper1}, the LM approach also produces comparable abundance estimates in the presence of larger grains. For instance, with grain sizes of $\sim$100 $\mu$m for H\textsubscript{2}O and H\textsubscript{2}SO\textsubscript{4}$\cdot$8H\textsubscript{2}O mixtures, the difference between LM and RT modeling can reach $\pm$15\% \citep{emran2025paper1}. In contrast, when $\sim$mm-sized grains are present, even in trace amounts as considered in this study, both LM and RT modeling predict comparable results (within $\pm$2\% difference), regardless of the mixing ratios. The diverse geology on the surface of Europa \citep[e.g.,][]{greeley2000geologic, prockter2009morphology, daubar2024planned, leonard2024global} is likely to host mixtures of small and coarse particles. Thus, both LM and RT modeling can be confidently applied to terrains host $\sim$mm-sized coarser H\textsubscript{2}O ice grains \citep[e.g.,][]{ligier2016vlt, king2022compositional}. Fig. \ref{fig5} represents geologic features associated with cryovolcanic activity and surface–subsurface exchange processes on Europa, where larger ($\sim$mm-sized) water-ice grains have been detected using near-infrared spectral data from the Galileo mission \citep{emran2025nh}. The linear crisscrossing features in the figure (Fig. \ref{fig5}) represent fractures within Europa’s ice shell that may facilitate material transport between the shallow subsurface and the surface. Ammonia-bearing coarse-grained H\textsubscript{2}O ice has been detected along these fractured regions using the NIR spectral modeling \citep{emran2025nh}.

Interestingly, across all mixing ratios and temperatures, the LM model produced lower RMSE values than the RT model, even though the RT-derived abundance estimates were slightly better in most of the mixing ratios. As an example, for the 77\% ($\sim$70 $\mu$m) : 23\% ($\sim$1060 $\mu$m) mixture, the RMSE values at 100K are 0.005 and 0.013 for LM and RT, respectively (Fig. \ref{fig2}), and at 120K, they are 0.006 and 0.014 (Fig. \ref{fig3}). Despite the higher RMSE, the RT model renders slightly more accurate abundance estimates than LM at this mixture ratio for both temperatures. In contrast, for the 27\% ($\sim$70 $\mu$m) : 73\% ($\sim$1060 $\mu$m) mixture at 100K (Fig. \ref{fig2}), the lower RMSE value for LM modeling (0.008) results in a slightly better abundance estimate than RT modeling (0.017). However, for the same mixture at 120K (Fig. \ref{fig3}), although LM modeling results in a lower RMSE (0.009) compared to RT (0.018), it does not correspond to a better abundance estimate. This suggest that spectral fit quality or goodness of fit— as quantified by metrics such as RMSE or chi-square ($\chi$\textsuperscript{2})— cannot be directly used to claim the accuracy of abundance estimates, consistent with \citet{emran2025paper1}. This lack of correlation between RMSE and abundance accuracy further emphasizes that a good spectral fit does not necessarily imply physically accurate parameter retrieval. Supporting that, \citet{khuller2025quantitative} also notes the inferred grain sizes in simplified Hapke models can deviate significantly from true constrained values even when the model spectral albedo closely reproduces the laboratory-measured albedo (or field measurement). This suggests that the apparent agreement between modeled and observed albedo may reflect parameter degeneracy inherent to the coupling of the phase function and scattering albedo, rather than a strictly physical validity of the retrieved parameters.

As observed in this study, the RMSE, which measures how well the synthetic spectrum from LM or RT approximates the laboratory spectrum at NIR wavelengths, does not consistently correlate with the difference between the calculated and laboratory abundances for LM and RT models (Figs. \ref{fig2}-\ref{fig3}). Thus, while defining a single global metric that combines both RMSE and abundance accuracy is not straightforward, integrating the results from existing studies suggests that RT modeling predicts consistently better abundance estimates across a wide range of compositions and grain sizes. A lower RMSE can sometimes indicate better numerical fitting of spectra without corresponding improvements in modeled physical parameters. Supporting this, a previous study has also shown that better spectral fits (i.e., lower RMSE) did not necessarily translate into more accurate estimates of grain size of H\textsubscript{2}O ice \citep{khuller2025quantitative}. Furthermore, when comparing LM and RT methods using Galileo/NIMS data on Europa, \cite{shirley2016europa} also reports that the goodness of fit (viz. $\chi$\textsuperscript{2}) does not necessarily correspond to an accurate representation of surface composition. Thus, this study cautions that, while RMSE or $\chi$\textsuperscript{2} values can indicate how well a model fits the observed data, they should not be solely relied upon as consistent indicators or global metrics of model performance for predicting physical parameters, such as abundance estimates on Europa.

\cite{emran2025paper1} showed that, for small grain sizes ($\sim$100 $\mu$m) in H\textsubscript{2}O ice and H\textsubscript{2}SO\textsubscript{4}$\cdot$8H\textsubscript{2}O mixtures, RT modeling provides much more accurate abundance estimates within $\pm$5\%, whereas LM estimates can be less accurate, with discrepancies up to $\pm$5-15\%. This difference arises because, when particle sizes are much smaller than the photon penetration depth (\textit{d} \(<\)\(<\) \textit{$\delta$}), multiple scattering effects  \citep[e.g,][]{hapke1981bidirectional, hapke2012theory} dominate in intimate mixtures. Multiple scattering can significantly alter the spectral characteristics of the ice, even when present in small amounts \citep[e.g.,][]{ hayes2025insights}. At near-infrared wavelengths, the photon penetration depth for H\textsubscript{2}O ice is on the order of hundreds of micrometers due to the low imaginary part (\textit{k}) of the optical constants \citep[e.g.,][]{mastrapa2008optical, mastrapa2009optical, grundy1998temperature}, as \textit{$\delta$} is inversely proportional to \textit{k} \citep[e.g.,][]{clark1984reflectance}. Radiative transfer–based intimate mixture modeling incorporates these multiple scattering effects (cf. Eq. \ref{eq:3}), capturing the complex scattering behavior of the mixture and resulting in accurate abundance estimates for small grains. In contrast, LM does not account for the complex interaction of light with granular media— ignores penetration depth and multiple scattering effect— leading to larger discrepancies in abundance estimates for small particles in H\textsubscript{2}O and H\textsubscript{2}SO\textsubscript{4}$\cdot$8H\textsubscript{2}O mixtures.

Nonetheless, when larger ($\sim$ mm-sized) grains are present, the photon penetration depth becomes comparable to the grain diameter (\textit{d} $\approx$ \textit{$\delta$}), and multiple scattering effects become less significant \citep[e.g.,][]{hapke1981bidirectional, clark1984reflectance}. Consequently, RT and LM modeling produce comparable abundance estimates, as LM modeling inherently does not account for multiple scattering. Thus, in mixtures containing larger H\textsubscript{2}O grains, even in small amounts, no significant difference is observed between the abundance estimates of H\textsubscript{2}O ice obtained from linear and intimate mixture modeling. This is further supported by the observation that, as the abundance of larger grains increases from 23\% to 73\% in the mixture, RT-derived estimates slightly deteriorate, leading to an average modeled abundance offset of $\pm 3$\% to $\pm 7$\% and $\pm 9$\% at 100 K (Fig. \ref{fig2}) and 120 K (Fig. \ref{fig3}), respectively, relative to the true laboratory values (Tables \ref{tab1} - \ref{tab2}). However, considering the 3\% uncertainty in the laboratory measurements \citep{stephan2021vis}, the RT-derived abundances remain reasonably accurate when smaller grains dominate the mixture. These observations are consistent with spectral mixture modeling of laboratory water ice and salt (epsomite) at cryogenic temperatures relevant to Europa, which likewise showed no discernible difference between linearly and intimately mixed spectral modeling \citep{berdis2025near}. Moreover, \cite{berdis2025near} also reported that the discrepancy between the modeled and laboratory spectra of H\textsubscript{2}O ice became more apparent with larger grain size in intimate mixture modeling. Thus, it is noteworthy that intimate mixture modeling — despite accounting for the non-linear scattering of photons among particles with differing transmission, refraction, and absorption properties and varied geometries \citep[e.g,][]{clark1984spectral} — did not result substantially more accurate estimates than linear modeling, consistent with both laboratory \citep{berdis2025near} and observational \citep{shirley2016europa} studies. Future investigations are warranted to compare both spectral modeling approaches using a more diverse set of materials—particularly mixtures involving strongly and weakly absorbing components—across a range of grain sizes and shapes.

The present work and previous studies \citep{emran2025paper1, berdis2025near} validated the linear and intimate mixture modeling using laboratory binary mixtures of H\textsubscript{2}O ice with H\textsubscript{2}SO\textsubscript{4}$\cdot$8H\textsubscript{2}O and hydrated sulfate salt epsomite. However, Europa’s surface hosts compositionally diverse mixtures containing species such as CO\textsubscript{2}, H\textsubscript{2}O\textsubscript{2}, SO\textsubscript{2}, NH\textsubscript{3}-bearing compounds, and various hydrated sulfate and chloride salts \citep[e.g.,][]{lane1981sulphur, carlson1999sulfuric, carlson1999b, carlson2009europa,shirley2010europa, brown2013salts,trumbo2019sodium,villanueva2023endogenous, becker2024exploring, emran2025nh}. Therefore, laboratory mixture spectra involving more than two components — together with a robust set of laboratory-derived optical constants — are warranted to assess spectral modeling performance comprehensively across Europa’s surface materials. This study also employed a simplified Hapke model configuration (Eq. \ref{eq:3}), assuming no backscattering, roughness, and porosity effects, and adopting an isotropic scattering function, appropriate for the laboratory measurement geometry (intermediate phase angles) and known particle sizes \citep{mustard1987quantitative, mustard1989photometric}. While imaging spectrometers onboard spacecraft often operate under similar geometries (phase angles of 15° – 40°), remote observations can involve different viewing conditions, and the grain sizes of surface materials may not always be well constrained or unknown. In such cases, it is imperative to properly account for Hapke parameters \citep{hapke1981bidirectional, hapke2012theory} describing backscattering, porosity, roughness, and the particle phase function in spectral modeling. Note that the Hapke parameters for Europa’s surface have been derived from photometric analyses \citep[e.g.,][]{buratti1985photometry, buratti1995photometry, domingue1992disk, belgacem2018estimation, belgacem2020regional} and are incorporated into radiative transfer modeling for remote sensing applications \citep[e.g.,][]{berdis2022europa, mermy2023selection}.

In summary, while \citet{emran2025paper1} shows that RT modeling performs significantly better for terrains composed of smaller grains ($\sim$100 $\mu$m), such as the leading hemisphere or near the poles of the trailing hemisphere of Europa \citep[e.g.,][]{kieffer1974frost, hansen2004amorphous, carlson2005distribution, carlson2009europa, moore2009surface, becker2024exploring}, the results of this study suggest that the presence of larger grains ($\sim$mm-sized) in the mixture leads to comparable abundance estimates from both linear mixture modeling and radiative transfer–based intimate mixture modeling using Hapke model \citep{hapke1981bidirectional}. Thus, incorporating the findings from existing studies, I conclude that RT modeling is the preferred approach for compositional analysis of Europa's surface at near-infrared wavelengths, regardless of material grain size, ice–non-ice compositional mixtures, or geological and regional variability. Nonetheless, the LM model can be reliably applied in regions where the surface mixture includes even trace amounts of $\sim$mm-sized particles, particularly in terrains shaped by surface–subsurface interactions, such as chaos regions and effusive cryovolcanic sites \citep[e.g.,][]{ligier2016vlt,king2022compositional, emran2025nh}. These findings provide a guideline for compositional analyses of Europa with future imaging spectrometers, including JUICE’s MAJIS \citep{poulet2024moons} and Europa Clipper’s MISE \citep{blaney2024mapping}, through the application of spectral mixture modeling.

\newpage
\section*{Nomenclature}
\textit{r} is reflectance or radiance factor

\textit{r\textsubscript{m} } is the reflectance of mixture

\textit{r\textsubscript{o} } is the bihemispherical reflectance for isotropic scatterers

\textit{f} is the fractional abundance/ factor

\textit{w} is the single scattering albedo (SSA)

\textit{w\textsubscript{m}} is the SSA of the mixture spectrum

\textit{i } is the incidence angle

\textit{e} is the emission angle

\textit{g} is the phase angle

\textit{B(g)} is the backscattering function

\textit{P(g)} is the single-particle phase function

\textit{H(x)} is the Chandrasekhar’s H functions

$\mu$\textsubscript{i} is the cosine of the incidence angle

$\mu$\textsubscript{e} is the cosine of the emission angle

 $\gamma$ is the albedo factor
 
 \textit{$\delta$} is the photon penetration depth

\textit{n} is the real part of optical constants

\textit{k} is the imaginary part of optical constants

\textit{d}  is the diameter or grain size

\textit{$\langle D \rangle$} is the "effective diameter"

\section*{Data Availability}
The laboratory spectral data used used in this study were collected from \cite{stephan2021vis} and can be accessed from SSHADE at \href{https://www.sshade.eu/}{https://www.sshade.eu/} \citep{schmitt2018solid}. Galileo Solid State Imaging (SSI) data used in this study can be found in the National Aeronautics and Space Administration's Planetary Data System: Imaging Node Server at \href{https://pdsimage2.wr.usgs.gov/}{https://pdsimage2.wr.usgs.gov/} \citep{malaska2024updated}.

\section*{Declaration of generative AI}
During the preparation of this work the author(s) used OpenAI in order to improve the readability and language of the manuscript. After using this tool/service, the author(s) reviewed and edited the content as needed and take(s) full responsibility for the content of the publication.

\section*{Acknowledgments}
This research was carried out at the Jet Propulsion Laboratory (JPL), California Institute of Technology, under a contract with the National Aeronautics and Space Administration (80NM0018D0004). I acknowledge JPL’s High-Performance Computing supercomputer facility, which was funded by JPL’s Information and Technology Solutions Directorate. I also thank Kathryn M. Stack for valuable discussions on spectral modeling and Aditya R. Khuller for insightful feedback on the manuscript. Copyright © 2025. California Institute of Technology. Government sponsorship acknowledged.
\bibliographystyle{elsarticle-harv} 
\bibliography{ref}
\end{document}